\documentclass[letterpaper, 10 pt, conference]{ieeeconf} 
\IEEEoverridecommandlockouts  
\usepackage{amsmath,amssymb,amsfonts}
\usepackage{algorithmic}
\usepackage{algorithm}
\usepackage{graphicx}
\usepackage{textcomp}
\usepackage{xcolor}
\usepackage{bm}
\usepackage{mathtools}
\usepackage{tikz}
\usepackage{cite}
\usepackage{booktabs}
\usepackage{newtxtext}

\usepackage{subcaption}
\usepackage{hyperref}
\usepackage{url}
\usepackage{xcolor}
\hypersetup{
    colorlinks=true,
    linkcolor=blue,
    citecolor=blue,
    urlcolor=blue
}

\newtheorem{proposition}{Proposition}
\newtheorem{remark}{Remark}

\begin{document}

\title{\LARGE \bf Task-Aware Datamodel-Based Column Selection for Nonlinear Data-Enabled Predictive Control}

\author{
Jiachen Li$^{1}$, Shihao Li$^{1}$, Jiamin Xu$^{1}$, Soovadeep Bakshi$^{1}$, and Dongmei Chen$^{1}$%
\thanks{$^{1}$All authors are with the Department of Mechanical Engineering, The University of Texas at Austin, Austin, TX 78712, USA.
        {\tt\small \{jiachenli, shihaoli01301, jiaminxu, soovadeepbakshi, dmchen\}@utexas.edu}}%
}

\maketitle

\begin{abstract}
Nonlinear Data-Enabled Predictive Control (DeePC) often requires online data reduction to stay computationally tractable. Recent methods choose Hankel columns based on geometric proximity to the current operating condition, but proximity is only an indirect proxy for what actually matters: control performance. In this work, we take a different view—treating each column as an individual data point whose inclusion or exclusion has a measurable effect on closed-loop cost—and draw on data attribution ideas to learn a \emph{data} model: a context-dependent linear surrogate that maps the current initial trajectory and reference to per-column influence scores, enabling fast online top-$K$ selection. We show that, under Bernoulli subset sampling, the learned coefficients have a natural interpretation as average marginal effects, and that top-$K$ selection is optimal for the surrogate under a cardinality constraint. We also provide a suboptimality bound that links surrogate accuracy to the quality of the selected subset. Experiments on ground vehicle and quadrotor benchmarks show that the proposed selector is competitive with and often outperforms existing baselines, while achieving up to $20\times$ computational savings over full-data DeePC.
\end{abstract}

\section{Introduction}

Model Predictive Control (MPC) is a widely used framework for constrained control, but it relies on having an accurate system model \cite{hewing2020learning}. When such a model is hard to obtain—whether because of complex nonlinear dynamics or limited first-principles knowledge—data--driven approaches that bypass explicit modeling and work directly from measured trajectories become appealing \cite{depersis2020formulas,rotulo2022online,hewing2020learning}.

Data-Enabled Predictive Control (DeePC) takes this idea further by replacing a parametric model entirely with a behavioral representation built from input-output data \cite{deepc,markovsky2021behavioral}. For linear time-invariant systems, Willems'fundamental lemma guaranties that every trajectory of a controllable system can be expressed using a single persistently exciting trajectory \cite{willems,vanwaarde2020informativity}, and regularized variants have since broadened DeePC's practical applicability to settings with noise and uncertainty \cite{coulson2019regularized,coulson2022distributionally,dorfler2023bridging,berberich2021datadriven}.

Nonlinear systems, however, remain a challenge. The exact trajectory-spanning property no longer holds, and the Hankel matrix acts only as an approximate local predictor \cite{berberich2021datadriven}. Achieving adequate coverage of the operating envelope in practice demands large datasets, which produce Hankel matrices with thousands of columns. Since computational cost grows with the number of retained columns, online implementation quickly becomes difficult when broad coverage is needed \cite{deepc,coulson2019regularized}. This tension between data richness and computational tractability has spurred a growing line of work on online data reduction for DeePC. One effective strategy is to select columns based on geometric proximity to the current operating condition, which can dramatically cut the computational burden while retaining locally relevant information \cite{choose_wisely}.

Geometric proximity, though, is only an indirect proxy for what ultimately matters: control performance. Two equally close subsets can lead to very different closed-loop behavior, and the same initial condition may call for different data depending on the reference trajectory being tracked. Recent work has begun to account for this reference dependence \cite{beerwerth2025less}, but the selection mechanism is still geometric rather than performance-driven and does not directly quantify how much a given column helps or hurts the controller's ability to track a specific target. This observation suggests a change in perspective, one inspired by \emph{data attribution} in machine learning, where each training example is assigned a score reflecting its contribution to a model's loss \cite{datamodels}: we can view each Hankel column as an individual data point whose inclusion or exclusion has a measurable influence on closed-loop performance. Building on this analogy, we learn offline a context-dependent linear surrogate that maps a binary column-selection vector to the realized short-horizon DeePC loss. The surrogate coefficients---which depend on the current initial trajectory and reference window---act as per-column influence scores, quantifying the expected marginal effect of including each column on tracking cost. At runtime, columns are simply ranked by these scores, and the top-$K$ most beneficial ones are retained, reducing the selection step to sorting a single vector with negligible overhead. The approach is an empirical surrogate-based strategy in the spirit of recent nonlinear DeePC work, where practical effectiveness is established experimentally \cite{choose_wisely,berberich2021datadriven,dorfler2023bridging}.

Our main contributions are as follows. We formulate a context-dependent linear surrogate that maps initial and reference trajectories to per-column influence scores, together with a consistent two-stage pipeline for offline training and online selection. We establish three surrogate-level properties: an average marginal-effect interpretation, top-$K$ optimality under a cardinality constraint, and a suboptimality bound. Experiments on ground vehicle and quadrotor benchmarks show that the proposed selector is competitive with and often outperforms existing baselines, while achieving up to $20\times$ computational savings over full-data DeePC.

The remainder of the paper is organized as follows. Section~II sets up the problem, introducing the Hankel matrix construction, the reduced DeePC formulation, and the learning objective. Section~III develops the context-dependent linear surrogate and establishes its theoretical properties. Section~IV describes the offline training pipeline and the online selection procedure. Section~V presents experimental results on both benchmarks, Section~VI discusses limitations and future directions, and Section~VII concludes the paper.

\section{Problem Setting}

\subsection{Nonlinear system and Hankel matrices}

Consider the discrete-time nonlinear system
\begin{align}
    x_{k+1} &= f(x_k,u_k), \label{eq:dynamics}\\
    y_k &= h(x_k), \label{eq:output}
\end{align}
where $x_k \in \mathbb{R}^n$, $u_k \in \mathbb{R}^m$, and $y_k \in \mathbb{R}^p$. The maps $f$ and $h$ are unknown, and only input-output data are available.

Given a pre-collected input-output trajectory $\{(u_k,y_k)\}_{k=0}^{T-1}$ and a depth
\begin{equation}
    L = T_{\mathrm{ini}} + N,
\end{equation}
define the Hankel matrices
\begin{equation}
    H_L(u)=
    \begin{bmatrix}
        u_0 & u_1 & \cdots & u_{T-L} \\
        u_1 & u_2 & \cdots & u_{T-L+1} \\
        \vdots & \vdots & \ddots & \vdots \\
        u_{L-1} & u_L & \cdots & u_{T-1}
    \end{bmatrix},
\end{equation}
and similarly $H_L(y)$. Let
\begin{equation}
    M = T - L + 1
\end{equation}
denote the number of columns. We partition
\begin{equation}
    \begin{bmatrix}U_p \\ U_f\end{bmatrix} = H_L(u), \qquad
    \begin{bmatrix}Y_p \\ Y_f\end{bmatrix} = H_L(y),
\end{equation}
where the first $T_{\mathrm{ini}}$ block rows are the past and the remaining $N$ block rows are the future.

For linear time-invariant systems and persistently exciting data, these matrices admit an exact behavioral interpretation \cite{willems,markovsky2021behavioral}. In the nonlinear setting considered here, they serve only as an approximate local predictor. Accordingly, the proposed column selector is not justified by nonlinear trajectory spanning; it is an empirical mechanism for improving reduced DeePC performance.

\subsection{Reduced DeePC with selected columns}

At time $t$, let
\begin{equation}
    u_{\mathrm{ini},t} \in \mathbb{R}^{mT_{\mathrm{ini}}}, \qquad
    y_{\mathrm{ini},t} \in \mathbb{R}^{pT_{\mathrm{ini}}}
\end{equation}
be the measured recent input-output trajectory, and let
\begin{equation}
    r_t = [r_t^\top \; r_{t+1}^\top \; \cdots \; r_{t+N-1}^\top]^\top \in \mathbb{R}^{pN}
\end{equation}
be the current reference window. As in DeePC, the initial trajectory is used as a practical finite-memory summary of the current operating condition. We do not assume that it is an exact state representation for the nonlinear system.

For a selected column set $\mathcal{S} \subseteq \{1,\dots,M\}$, denote the reduced Hankel blocks by
\begin{equation}
    U_p^{\mathcal{S}},\; Y_p^{\mathcal{S}},\; U_f^{\mathcal{S}},\; Y_f^{\mathcal{S}}.
\end{equation}
The corresponding reduced DeePC problem is
\begin{align}
    \min_{g,u_f,y_f,\sigma_y}\quad &
    \sum_{k=0}^{N-1}
    \left(
        \|y_{f,k}-r_{t+k}\|_Q^2 + \|u_{f,k}\|_R^2
    \right)
    \nonumber\\
    &\quad + \lambda_g \|g\|_2^2 + \lambda_y \|\sigma_y\|_2^2
    \label{eq:deepc_obj}\\
    \text{s.t.}\quad &
    \begin{bmatrix}
        U_p^{\mathcal{S}} \\
        Y_p^{\mathcal{S}} \\
        U_f^{\mathcal{S}} \\
        Y_f^{\mathcal{S}}
    \end{bmatrix}
    g
    =
    \begin{bmatrix}
        u_{\mathrm{ini},t} \\
        y_{\mathrm{ini},t} + \sigma_y \\
        u_f \\
        y_f
    \end{bmatrix},
    \label{eq:deepc_const}\\
    & u_f \in \mathcal{U}, \qquad y_f \in \mathcal{Y}. \nonumber
\end{align}
Here $g \in \mathbb{R}^{|\mathcal{S}|}$ is the trajectory coefficient vector, $Q \succeq 0$ and $R \succ 0$ are tracking and input cost matrices, $\lambda_g > 0$ and $\lambda_y > 0$ are regularization weights, and $\sigma_y$ is a slack variable that accommodates noise-induced mismatch in the initial output trajectory. The constraint sets $\mathcal{U}$ and $\mathcal{Y}$ encode input and output bounds.

When $|\mathcal{S}| \ll M$, the reduced problem is much smaller than the full DeePC. The challenge is therefore to choose a subset that preserves good control performance.

\begin{remark}
Prior work in nonlinear DeePC suggests that a smaller, carefully selected subset of columns can sometimes outperform the full dataset in tracking tasks \cite{choose_wisely}. A plausible explanation is that, in the nonlinear setting, the regularized DeePC predictor is only approximate, so columns that are less relevant to the current operating condition may worsen conditioning and degrade prediction quality. From this perspective, restricting attention to a locally relevant subset can be viewed as a form of local data selection.
\end{remark}

\subsection{Learning objective}

Let $s \in \{0,1\}^M$ denote the subset indicator, where $s_j=1$ iff column $j$ is selected. Let
\begin{equation}
    c_t := (u_{\mathrm{ini},t}, y_{\mathrm{ini},t}, r_t)
\end{equation}
denote the control context.

To define a training target that matches online deployment, fix a selection-update horizon $H_{\mathrm{sel}} \ge 1$. For a context $c$ and a subset $s$, let
\begin{equation}
    J_H(s;c)
\end{equation}
denote the realized cost over the next $H_{\mathrm{sel}}$ plant steps obtained by solving \eqref{eq:deepc_obj}--\eqref{eq:deepc_const} with subset $s$ and keeping that subset fixed over those $H_{\mathrm{sel}}$ steps. More explicitly,
\begin{equation}
    J_H(s;c)
    :=
    \sum_{\tau=0}^{H_{\mathrm{sel}}-1}
    \left(
        \|y_{t+\tau}-r_{t+\tau}\|_Q^2 +
        \|u_{t+\tau}\|_R^2
    \right),
    \label{eq:Jh_def}
\end{equation}
where the trajectory $(u_{t+\tau}, y_{t+\tau})$ is generated by the plant under the reduced DeePC controller associated with $s$.

The purpose of the proposed datamodel is to approximate the map
\begin{equation}
    (s,c) \mapsto J_H(s;c)
\end{equation}
well enough that columns can be ranked online without solving many reduced DeePC problems.

\section{Context-Dependent Linear Surrogate}

\subsection{Surrogate model}

We model the subset-to-performance map using the context-dependent linear surrogate
\begin{equation}
    \widehat{J}_H(s;c) = s^\top \theta(c) + \theta_0(c),
    \label{eq:surrogate}
\end{equation}
where
\begin{equation}
    \theta(c) \in \mathbb{R}^M, \qquad \theta_0(c) \in \mathbb{R}
\end{equation}
are context-dependent coefficients. The model is linear in the subset indicator $s$ but nonlinear in the context through the map $c \mapsto (\theta(c),\theta_0(c))$.

We do not expect this surrogate to be exact. Its purpose is more modest: to give us a tractable way of scoring columns by how much they are likely to help (or hurt) the short-horizon DeePC loss. The additive form is a deliberate simplification—it cannot capture interactions between pairs of columns, but in return, it is easy to interpret and leads directly to the top-$K$ selection rule we derive next.

\subsection{Population interpretation under Bernoulli subset sampling}

The following proposition clarifies what the coefficients mean when the surrogate is fitted under independent Bernoulli subset sampling.

\begin{proposition}[Average marginal effect]
Fix a context $c$ and assume $J_H(s;c)\in L^2$. Let $s_1,\dots,s_M$ be i.i.d.\ $\mathrm{Bernoulli}(\alpha)$ with $\alpha\in(0,1)$, and let $(\theta_0^\star(c),\theta^\star(c))$ solve
\begin{equation}
    \min_{\theta_0,\theta}
    \ \mathbb{E}\!\left[
        \left(
            J_H(s;c)-\theta_0-\theta^\top s
        \right)^2
        \,\middle|\, c
    \right].
    \label{eq:pop_ls}
\end{equation}
Then, for each $j\in\{1,\dots,M\}$,
\begin{equation}
    \theta_j^\star(c)
    =
    \mathbb{E}[J_H(s;c)\mid s_j=1,c]
    -
    \mathbb{E}[J_H(s;c)\mid s_j=0,c].
    \label{eq:marginal_effect}
\end{equation}
\end{proposition}

\begin{proof}
Because an intercept is included and the regressors are independent, the normal equations for \eqref{eq:pop_ls} yield
\begin{equation}
    \mathrm{Cov}(s_j, J_H(s;c)\mid c)
    =
    \theta_j^\star(c)\,\mathrm{Var}(s_j\mid c).
    \label{eq:cov_eq}
\end{equation}
Since $s_j \sim \mathrm{Bernoulli}(\alpha)$,
\begin{equation}
    \mathrm{Var}(s_j\mid c) = \alpha(1-\alpha).
\end{equation}
Moreover,
\begin{equation}
\begin{aligned}
\mathrm{Cov}(s_j,J_H(s;c)\mid c)
&= \mathbb{E}[s_j J_H(s;c)\mid c] \\
&\quad - \alpha\,\mathbb{E}[J_H(s;c)\mid c] \\
&= \alpha\,\mathbb{E}[J_H(s;c)\mid s_j=1,c] \\
&\quad - \alpha\,\mathbb{E}[J_H(s;c)\mid c].
\end{aligned}
\end{equation}
Using the law of total expectation,
\begin{align}
    \mathbb{E}[J_H(s;c)\mid c]
    &=
    \alpha\,\mathbb{E}[J_H(s;c)\mid s_j=1,c]
    \nonumber\\
    &\quad +
    (1-\alpha)\,\mathbb{E}[J_H(s;c)\mid s_j=0,c],
\end{align}
we obtain
\begin{equation}
\begin{aligned}
\mathrm{Cov}(s_j,J_H(s;c)\mid c)
&= \alpha(1-\alpha) \Bigl(
   \mathbb{E}[J_H(s;c)\mid s_j=1,c] \\
&\qquad\qquad
   - \mathbb{E}[J_H(s;c)\mid s_j=0,c]
   \Bigr).
\end{aligned}
\end{equation}
Substituting into \eqref{eq:cov_eq} proves \eqref{eq:marginal_effect}.
\end{proof}

\begin{remark}
Proposition~1 tells us that, under Bernoulli subset sampling, the population coefficient $\theta_j^\star(c)$ captures the average effect of toggling column $j$ on and off, where the average is taken over all $2^{M-1}$ configurations of the remaining columns, weighted by the Bernoulli measure. This is not an exact decomposition of nonlinear DeePC performance—it is a linear summary—and the coefficients will generally depend on the inclusion probability $\alpha$ used during training.
\end{remark}
\subsection{Surrogate-optimal top-\texorpdfstring{$K$}{K} selection}

Given a subset budget $K$, the learned surrogate yields a simple ranking rule.

\begin{proposition}[Top-$K$ optimality]
Fix a context $c$ and a budget $K\in\{1,\dots,M\}$. Any solution of
\begin{equation}
    \min_{s\in\{0,1\}^M}
    \ \widehat{J}_H(s;c)
    \quad\text{s.t.}\quad
    \mathbf{1}^\top s = K
    \label{eq:card_surrogate}
\end{equation}
is obtained by setting $s_j=1$ for the $K$ smallest components of $\theta(c)$.
\end{proposition}

\begin{proof}
Under the constraint $\mathbf{1}^\top s = K$, the bias $\theta_0(c)$ is constant. Therefore, \eqref{eq:card_surrogate} reduces to minimizing
$\sum_{j=1}^M s_j \theta_j(c)$
over binary vectors with exactly $K$ ones. Any minimizer selects the $K$ smallest coefficients.
\end{proof}

\begin{remark}
Proposition~2 concerns only the learned surrogate \eqref{eq:surrogate}; it does not claim that the resulting subset is globally optimal for the true realized loss $J_H(\cdot;c)$. When the true loss involves significant column interactions, the additive surrogate may misrank some subsets.
\end{remark}

\subsection{Surrogate-suboptimality bound}

The next proposition quantifies how surrogate approximation error transfers to subset quality under a cardinality constraint.

\begin{proposition}[Suboptimality bound]
Fix a context $c$ and a budget $K$. Suppose that
\begin{equation}
    \sup_{\substack{s\in\{0,1\}^M\\ \mathbf{1}^\top s = K}}
    \left|
        J_H(s;c) - \widehat{J}_H(s;c)
    \right|
    \le \varepsilon(c).
    \label{eq:uniform_err}
\end{equation}
Let $s^\star(c)$ and $\widehat{s}(c)$ be minimizers of $J_H(\cdot;c)$ and $\widehat{J}_H(\cdot;c)$, respectively, over $\{s\in\{0,1\}^M : \mathbf{1}^\top s = K\}$. Then
\begin{equation}
    J_H(\widehat{s}(c);c) - J_H(s^\star(c);c)
    \le 2\varepsilon(c).
    \label{eq:2eps}
\end{equation}
\end{proposition}

\begin{proof}
By \eqref{eq:uniform_err},
$J_H(\widehat{s};c)
\le
\widehat{J}_H(\widehat{s};c) + \varepsilon(c)$.
Because $\widehat{s}$ minimizes the surrogate,
$\widehat{J}_H(\widehat{s};c)
\le
\widehat{J}_H(s^\star;c)$.
Applying \eqref{eq:uniform_err} again gives
$\widehat{J}_H(s^\star;c)
\le
J_H(s^\star;c) + \varepsilon(c)$.
Combining the three inequalities yields \eqref{eq:2eps}.
\end{proof}

\begin{remark}
The bound in Proposition~3 is generic—it holds for any surrogate satisfying \eqref{eq:uniform_err} and has nothing to do with the additive structure of \eqref{eq:surrogate}. How useful it is in practice comes down to how small $\varepsilon(c)$ turns out to be, which is ultimately an empirical question that we leave for future investigation.
\end{remark}

\subsection{Context parameterization}

We parameterize the context-to-score map using a neural network
\begin{equation}
    g_\phi(c)
    =
    [\theta(c)^\top \; \theta_0(c)]^\top
    \in \mathbb{R}^{M+1},
\end{equation}
where the context vector is
\begin{equation}
    c = [u_{\mathrm{ini}}^\top \; y_{\mathrm{ini}}^\top \; r^\top]^\top
    \in \mathbb{R}^{(m+p)T_{\mathrm{ini}} + pN}.
\end{equation}
In the experiments, $g_\phi$ is implemented as a multilayer perceptron (MLP).

\section{Offline Learning and Online Selection}

\begin{figure}[t]
    \centering
    \includegraphics[width=\columnwidth]{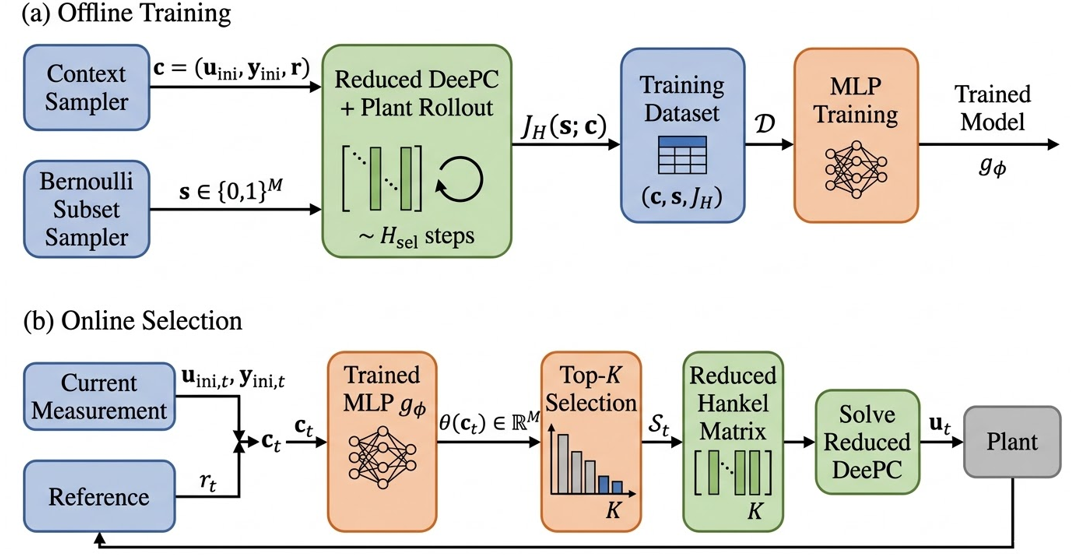}
    \caption{Datamodel-Based DeePC overview.}
    \label{fig:method_overview}
\end{figure}

Fig.~\ref{fig:method_overview} summarizes the two-stage pipeline. The offline stage (a) generates training data by rolling out reduced DeePC under random subsets, then trains a context-to-score network. The online stage (b) uses this network to select columns before solving the reduced controller.

\subsection{Offline data generation and training}

The offline stage generates training pairs of contexts, subsets, and realized losses according to the same reduced DeePC formulation and the same selection-update horizon used online. This alignment is important: the learned surrogate should predict the quantity that the controller actually cares about.

Let
$\widetilde{s} := [s^\top \; 1]^\top \in \mathbb{R}^{M+1}$
denote the augmented indicator, and let
$z_\phi(c) := [\theta(c)^\top \; \theta_0(c)]^\top \in \mathbb{R}^{M+1}$
be the network output. Given a training set
\begin{equation}
    \mathcal{D}
    =
    \{(c^{(i)}, s^{(i)}, J_H^{(i)})\}_{i=1}^{N_{\mathrm{train}}},
\end{equation}
we train $g_\phi$ by minimizing
\begin{equation}
    \mathcal{L}(\phi)
    =
    \frac{1}{N_{\mathrm{train}}}
    \sum_{i=1}^{N_{\mathrm{train}}}
    \left(
        \widetilde{s}^{(i)\top} z_\phi(c^{(i)})
        -
        J_H^{(i)}
    \right)^2
    +
    \lambda_\phi \|\phi\|_2^2,
    \label{eq:loss}
\end{equation}
where $\lambda_\phi > 0$ is a weight-decay coefficient.

The full offline procedure is summarized in Algorithm~\ref{alg:offline}.

\begin{algorithm}[t]
\caption{Offline Data Generation and Datamodel Training}
\label{alg:offline}
\begin{algorithmic}[1]
\REQUIRE Full Hankel blocks $(U_p,Y_p,U_f,Y_f)$, DeePC parameters $(Q,R,\lambda_g,\lambda_y,\mathcal{U},\mathcal{Y})$, context sampler $p(c)$, Bernoulli parameter $\alpha$, update horizon $H_{\mathrm{sel}}$, number of samples $N_{\mathrm{train}}$
\ENSURE Trained network $g_\phi$
\STATE Initialize dataset $\mathcal{D}\leftarrow\emptyset$
\FOR{$i=1$ to $N_{\mathrm{train}}$}
    \STATE Sample context $c^{(i)}=(u_{\mathrm{ini}}^{(i)}, y_{\mathrm{ini}}^{(i)}, r^{(i)}) \sim p(c)$
    \STATE Sample $s^{(i)}\in\{0,1\}^M$ with $s_j^{(i)}\sim \mathrm{Bernoulli}(\alpha)$ independently
    \STATE Let $\mathcal{S}^{(i)}=\{j:s_j^{(i)}=1\}$
    \STATE Solve reduced DeePC \eqref{eq:deepc_obj}--\eqref{eq:deepc_const} using columns $\mathcal{S}^{(i)}$
    \STATE Apply the resulting controller to the plant for $H_{\mathrm{sel}}$ steps while keeping $\mathcal{S}^{(i)}$ fixed
    \STATE Record $J_H^{(i)}$ via \eqref{eq:Jh_def}
    \STATE Append $(c^{(i)}, s^{(i)}, J_H^{(i)})$ to $\mathcal{D}$
\ENDFOR
\STATE Train $g_\phi$ by minimizing \eqref{eq:loss} over $\mathcal{D}$
\end{algorithmic}
\end{algorithm}

\begin{remark}
We use independent Bernoulli subset sampling in Algorithm~\ref{alg:offline} for two reasons: it exposes the surrogate to a wide variety of column combinations, and it is directly compatible with the marginal-effect interpretation established in Proposition~1. Since each training sample corresponds to a full rollout under a single fixed random subset, the total number of context--subset pairs seen during training depends on both $N_{\mathrm{train}}$ and the length of each rollout.
\end{remark}

\subsection{Online datamodel-based selection for DeePC}

Online, the learned network produces context-dependent column scores, and the DeePC controller is solved on the $K$ columns with the smallest predicted coefficients. The complete online procedure is given in Algorithm~\ref{alg:online}.

\begin{algorithm}[t]
\caption{Online Datamodel-Based Selection for DeePC}
\label{alg:online}
\begin{algorithmic}[1]
\REQUIRE Trained network $g_\phi$, full Hankel blocks $(U_p,Y_p,U_f,Y_f)$, subset budget $K$, DeePC parameters $(Q,R,\lambda_g,\lambda_y,\mathcal{U},\mathcal{Y})$
\FOR{each time step $t$}
    \STATE Measure the current initial trajectory $(u_{\mathrm{ini},t},y_{\mathrm{ini},t})$
    \STATE Form the current reference window $r_t$
    \STATE Form the context $c_t=(u_{\mathrm{ini},t},y_{\mathrm{ini},t},r_t)$
    \STATE Compute $[\theta(c_t)^\top \; \theta_0(c_t)]^\top = g_\phi(c_t)$
    \STATE Select the subset $\mathcal{S}_t$ corresponding to the $K$ smallest components of $\theta(c_t)$
    \STATE Construct reduced Hankel blocks $U_p^{\mathcal{S}_t}, Y_p^{\mathcal{S}_t}, U_f^{\mathcal{S}_t}, Y_f^{\mathcal{S}_t}$
    \STATE Solve reduced DeePC \eqref{eq:deepc_obj}--\eqref{eq:deepc_const}
    \STATE Apply the first control input $u_t = u_{f,0}^\star$
\ENDFOR
\end{algorithmic}
\end{algorithm}

\subsection{Computational cost}

The proposed controller adds an MLP forward pass and a ranking step before solving reduced DeePC. For an MLP with hidden width $d_h$ and $\ell$ layers, the forward-pass cost is $\mathcal{O}(\ell d_h^2 + d_h M)$. Extracting the top-$K$ indices requires $\mathcal{O}(M)$ expected time via linear-time selection. The main computational benefit comes from replacing the original DeePC problem with one whose coefficient vector $g$ has dimension $K$ instead of $M$, yielding a QP whose size scales with $K$ rather than $M$.

\section{Experimental Evaluation}

\subsection{Setup}

We evaluate the proposed selector on two nonlinear benchmarks: a ground vehicle tracking a raceline and a quadrotor tracking a three-dimensional figure-8 trajectory \cite{beerwerth2025less}. In all experiments, the downstream controller is the same regularized DeePC pipeline; only the online data-selection rule is changed. All methods share identical DeePC hyperparameters $(Q, R, \lambda_g, \lambda_y)$ and constraint sets.

The \emph{vehicle benchmark} uses the F1TENTH-style simulator on the S\~ao Paulo track with kinematic output
$y_t = [x_t \;\; y_t \;\; v_t \;\; \psi_t]^\top \in \mathbb{R}^4$
(position, velocity, heading) and control input
$u_t = [a_t \;\; \delta_t]^\top \in \mathbb{R}^2$
(acceleration, steering). The DeePC controller uses $T_{\mathrm{ini}}=5$, $N=10$. Each rollout lasts $60$\,s at $100$\,Hz with the planner updated at $10$\,Hz. Sensor noise is enabled ($0.05$\,m position, $0.05$\,m/s velocity, $0.01$\,rad heading). After preprocessing, the Hankel matrix contains $M=1{,}185$ columns.

The \emph{UAV benchmark} uses PyBullet with a Crazyflie-type vehicle tracking a figure-8 reference, with position output
$y_t = [x_t \;\; y_t \;\; z_t]^\top \in \mathbb{R}^3$
and four-dimensional attitude/thrust input. The controller uses $T_{\mathrm{ini}}=5$, $N=15$. Each rollout lasts $20$\,s at $250$\,Hz with the planner at $25$\,Hz. We test both \emph{nominal} (no noise/wind) and \emph{disturbed} (sensor noise and wind perturbations at default levels) conditions. The Hankel matrix contains $M=7{,}480$ columns.

We compare the proposed \emph{datamodel} selector against: \emph{contextual} (nearest-neighbor column selection in input-output feature space \cite{ beerwerth2025less}), \emph{select-DPC} (a geometry-based selector \cite{choose_wisely}), \emph{random} (uniform random selection), and \emph{full} (all $M$ columns). Subset budgets are $K \in \{30,60,90\}$ for the vehicle and $K \in \{40,70,100\}$ for the UAV. Each configuration is repeated over five seeds.

The context-to-score map $g_\phi$ is an MLP with three hidden layers of width $128$, trained with Adam \cite{adam} (learning rate $10^{-3}$, weight decay $10^{-6}$, batch size $32$, $100$ epochs). Each model is trained on $32$ rollouts generated by Bernoulli subset sampling with $\alpha = K/M$. Separate models are trained per benchmark and budget. We report weighted root-mean-square tracking error (wRMSE, mean\,$\pm$\,std over five seeds) and mean per-step computation time.

\subsection{Results}

Table~\ref{tab:combined_results} summarizes the closed-loop results. We highlight the lowest wRMSE among reduced-data methods in each row.

\begin{table*}[t]
\centering
\caption{Closed-loop results on vehicle and UAV benchmarks. Each entry reports mean\,$\pm$\,std of wRMSE over five seeds, followed by mean computation time\,[s]. }
\label{tab:combined_results}
\setlength{\tabcolsep}{3pt}
\renewcommand{\arraystretch}{1.15}
\scriptsize
\begin{tabular}{@{}llccccc@{}}
\toprule
 & $K$ & Datamodel & Contextual & Select-DPC & Random & Full \\
\midrule
\multicolumn{7}{c}{\textit{Vehicle} ($M=1{,}185$)} \\
\midrule
 & 30  & $0.270{\scriptstyle\pm0.031}$\,/\,0.018 & $0.316{\scriptstyle\pm0.028}$\,/\,0.009 & $\mathbf{0.265}{\scriptstyle\pm0.022}$\,/\,0.016 & $3.040{\scriptstyle\pm1.21}$\,/\,0.008 & --- \\
 & 60  & $\mathbf{0.172}{\scriptstyle\pm0.019}$\,/\,0.021 & $0.219{\scriptstyle\pm0.015}$\,/\,0.011 & $0.208{\scriptstyle\pm0.018}$\,/\,0.021 & $0.392{\scriptstyle\pm0.087}$\,/\,0.009 & --- \\
 & 90  & $\mathbf{0.122}{\scriptstyle\pm0.014}$\,/\,0.030 & $0.199{\scriptstyle\pm0.012}$\,/\,0.015 & $0.177{\scriptstyle\pm0.016}$\,/\,0.025 & $0.259{\scriptstyle\pm0.041}$\,/\,0.012 & --- \\
 & 1185 & --- & --- & --- & --- & $0.195{\scriptstyle\pm0.011}$\,/\,0.107 \\
\midrule
\multicolumn{7}{c}{\textit{UAV---nominal}} \\
\midrule
 & 40  & $\mathbf{0.152}{\scriptstyle\pm0.024}$\,/\,0.020 & $0.239{\scriptstyle\pm0.035}$\,/\,0.017 & $0.155{\scriptstyle\pm0.021}$\,/\,0.020 & $1.689{\scriptstyle\pm0.82}$\,/\,0.008 & --- \\
 & 70  & $0.051{\scriptstyle\pm0.008}$\,/\,0.024 & $\mathbf{0.045}{\scriptstyle\pm0.006}$\,/\,0.020 & $0.098{\scriptstyle\pm0.015}$\,/\,0.026 & $0.267{\scriptstyle\pm0.053}$\,/\,0.010 & --- \\
 & 100 & $\mathbf{0.019}{\scriptstyle\pm0.003}$\,/\,0.029 & $0.031{\scriptstyle\pm0.005}$\,/\,0.023 & $0.049{\scriptstyle\pm0.021}$\,/\,0.033 & $0.712{\scriptstyle\pm0.15}$\,/\,0.011 & --- \\
 & 7480 & --- & --- & --- & --- & $0.017{\scriptstyle\pm0.002}$\,/\,0.580 \\
\midrule
\multicolumn{7}{c}{\textit{UAV---disturbed}} \\
\midrule
 & 40  & $0.225{\scriptstyle\pm0.038}$\,/\,0.020 & $0.243{\scriptstyle\pm0.032}$\,/\,0.016 & $\mathbf{0.156}{\scriptstyle\pm0.019}$\,/\,0.020 & $2.015{\scriptstyle\pm0.94}$\,/\,0.009 & --- \\
 & 70  & $\mathbf{0.041}{\scriptstyle\pm0.007}$\,/\,0.025 & $0.047{\scriptstyle\pm0.006}$\,/\,0.019 & $0.094{\scriptstyle\pm0.013}$\,/\,0.027 & $0.267{\scriptstyle\pm0.048}$\,/\,0.010 & --- \\
 & 100 & $\mathbf{0.030}{\scriptstyle\pm0.005}$\,/\,0.029 & $0.035{\scriptstyle\pm0.004}$\,/\,0.022 & $0.805{\scriptstyle\pm0.017}/\,0.022$ & $0.680{\scriptstyle\pm0.13}$\,/\,0.011 & --- \\
 & 7480 & --- & --- & --- & --- & $0.020{\scriptstyle\pm0.002}$\,/\,0.481 \\
\bottomrule
\end{tabular}
\end{table*}

On the vehicle benchmark, select-DPC has the best performance at the smallest budget, but the data model becomes the best reduced-data method at the two larger budgets, where it substantially outperforms both geometry-based selectors. Notably, at $K=90$, the datamodel surpasses the full-data baseline, suggesting that restricting the predictor to a smaller, more relevant subset of data may improve performance in this nonlinear DeePC setting.

On the UAV nominal benchmark, all three structured selectors perform comparably at the smallest budget. At $K=100$, the datamodel nearly matches the full-data baseline while requiring roughly $20\times$ less computation.  The datamodel is best at the two larger budgets under both nominal and disturbed conditions.

Overall, the datamodel gives the best results in six out of nine test cases, especially when more columns are used. The geometry-based methods do well when very few columns are selected.

\begin{figure}[t]
    \centering
    \includegraphics[width=0.5\textwidth]{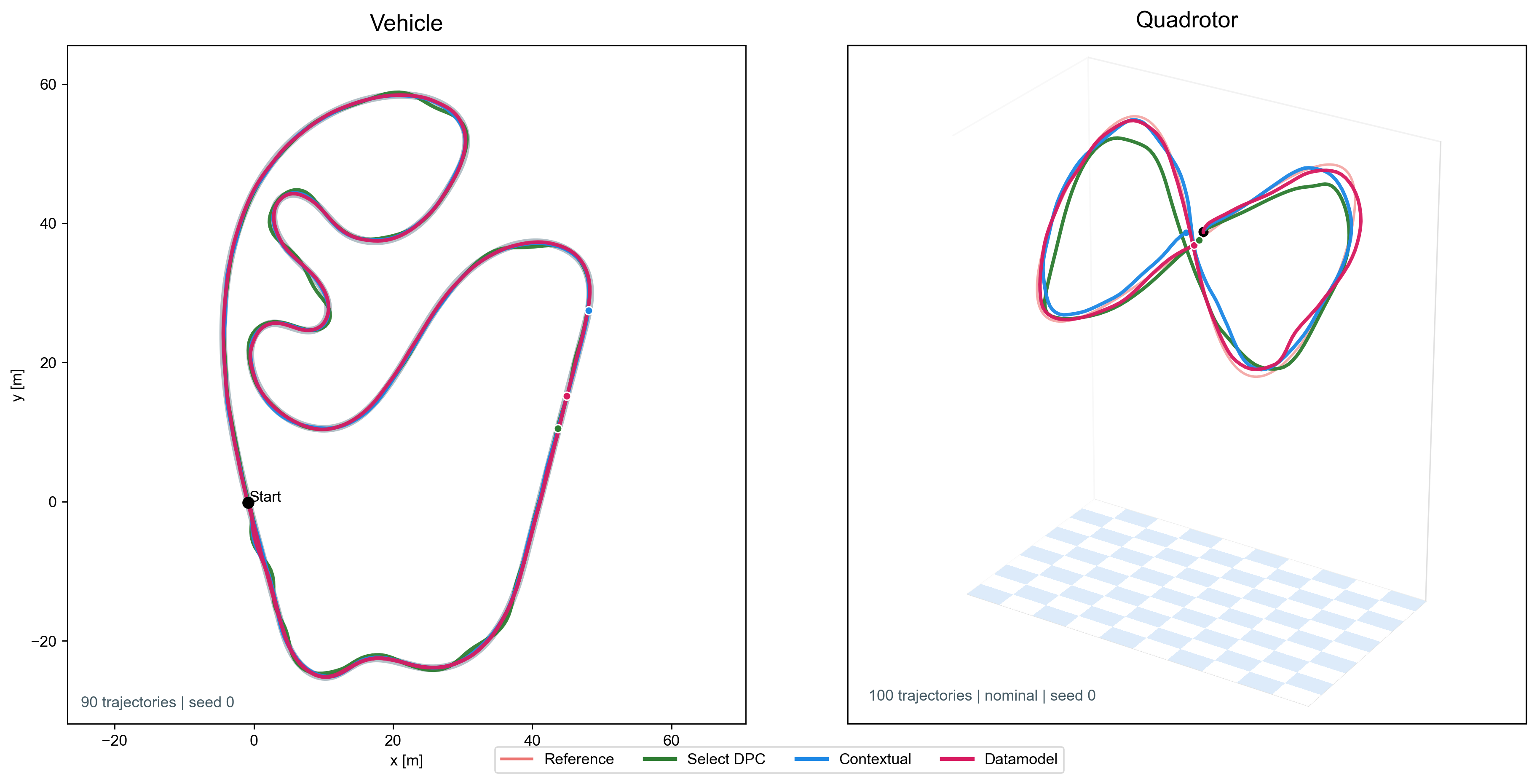}
    \caption{Closed-loop trajectories for the vehicle ($K{=}90$, left) and quadrotor ($K{=}100$, nominal, right). }
    \label{fig:trajectory_comparison}
\end{figure}

\subsection{Surrogate quality and sample efficiency}

To assess the additive surrogate's ranking ability, we evaluate the surrogate-predicted cost $\widehat{J}_H(s;c)$ and the true cost $J_H(s;c)$ on 100 random subsets for each of the 10 contexts sampled from the vehicle evaluation distribution. The mean Spearman rank correlation is $0.71 \pm 0.08$, and the surrogate-selected top-$K$ subset falls in the top 15th percentile of true cost in 8 of the 10 contexts (top 25th in all 10), confirming that the additive approximation produces near-best subsets.

To verify robustness to training set size, we train vehicle models at $K=60$ with $N_{\mathrm{train}} \in \{8, 16, 32, 64\}$ rollouts, obtaining wRMSE values of $0.241$, $0.198$, $0.172$, and $0.158$, respectively. Even with 16 rollouts, the datamodel (0.198) outperforms the contextual model (0.219), and performance improves monotonically with more data.

In terms of computation, at the largest tested budget, the datamodel achieves tracking accuracy within $1.1$--$1.6\times$ of the full-data baseline while requiring $4$--$20\times$ less computation, depending on the benchmark.

\section{Future Directions}

\textbf{Distribution shift and adaptive subset budget.}
The datamodel is trained and evaluated on contexts drawn from the same reference and disturbance distribution, so how well it transfers to substantially different reference families is an open question. For large distribution shifts, retraining is likely unavoidable. A natural next step is to explore domain-adaptation or meta-learning strategies that allow the surrogate to generalize across tasks without starting from scratch each time. Also, the column budget $K$ is currently fixed before deployment, but different operating conditions may call for more or fewer columns. Learning to adjust $K$ online could improve the computation--performance trade-off without manual tuning.

\textbf{Online adaptation.}
As it stands, the surrogate is trained entirely offline and remains fixed once deployed. If the system drifts or encounters conditions that were underrepresented in the training data, the predicted scores may become less reliable. Updating the surrogate online using closed-loop feedback—for instance, by fine-tuning on recently observed context—subset—cost tuples—could improve selection quality over time and reduce sensitivity to the distribution-shift issue mentioned above.

\textbf{Closed-loop stability guaranties.}
Formal closed-loop stability guaranties for reduced DeePC with dynamic column selection remain an open problem. Pairing the surrogate-based selector with robust DeePC formulations \cite{coulson2022distributionally,berberich2021datadriven} is a promising direction.

\section{Conclusion}

We presented a task-aware datamodel-based method for column selection in nonlinear DeePC. A context-dependent linear surrogate is learned offline to predict short-horizon DeePC performance, and its coefficients are used online to rank Hankel columns before solving a reduced controller. Experiments on ground vehicle and quadrotor benchmarks show the proposed selector achieves the best tracking accuracy in six of nine configurations while maintaining up to $20\times$ computational savings over full-data DeePC. 

\section*{Acknowledgment}
The authors thank Jaap Eising and Joshua N\"af for valuable discussions. Our experiments build on the codebase from \cite{beerwerth2025less}. Claude was used to assist with the language editing of this manuscript.

\bibliographystyle{IEEEtran}
\bibliography{references}

\end{document}